# Novel BCD Adders and Their Reversible Logic Implementation for IEEE 754r Format

Himanshu Thapliyal, Saurabh Kotiyal and M.B Srinivas
*Center for VLSI and Embedded System Technologies,*
*International Institute of Information Technology, Hyderabad-500019, India*
*\*Department of Computer Engineering, SIT, Kukas, Jaipur, India*
(thapliyalhimanshu@yahoo.com, saurabhkotiyal@yahoo.com, srinivas@iiit.net)

**Abstract**

*IEEE 754r is the ongoing revision to the IEEE 754 floating point standard and a major enhancement to the standard is the addition of decimal format. This paper proposes two novel BCD adders called carry skip and carry look-ahead BCD adders respectively. Furthermore, in the recent years, reversible logic has emerged as a promising technology having its applications in low power CMOS, quantum computing, nanotechnology, and optical computing. It is not possible to realize quantum computing without reversible logic. Thus, this paper also paper provides the reversible logic implementation of the conventional BCD adder as the well as the proposed Carry Skip BCD adder using a recently proposed TSG gate. Furthermore, a new reversible gate called TS-3 is also being proposed and it has been shown that the proposed reversible logic implementation of the BCD Adders is much better compared to recently proposed one, in terms of number of reversible gates used and garbage outputs produced. The reversible BCD circuits designed and proposed here form the basis of the decimal ALU of a primitive quantum CPU.*

## 1. Introduction

Nowadays, the decimal arithmetic is receiving significant attention as the financial, commercial, and Internet-based applications cannot tolerate errors generated by conversion between decimal and binary formats. Furthermore, a number of decimal numbers, such as 0.110, cannot be exactly represented in binary, thus, these applications often store data in decimal format and process data using decimal arithmetic software [1]. The advantage of decimal arithmetic in eliminating conversion errors also comes with a drawback; it is typically 100 to 1,000 times slower than binary arithmetic implemented in hardware.

Since, the decimal arithmetic is getting significant attention; specifications for it have recently been added to the draft revision of the IEEE 754 standard for Floating-Point Arithmetic. **IEEE 754r** is an ongoing revision to the IEEE 754 floating point standard [2,3]. Some of the major enhancements so far incorporated are the addition of 128-bit and decimal formats. Furthermore, three new decimal formats are described, matching the lengths of the binary formats. These have led to the decimal formats with 7, 16, and 34-digit significands, which may be normalized or unnormalized. In the proposed IEEE 754r format, for maximum range and precision, the formats merge part of the exponent and significand into a *combination field*, and compress the remainder of the significand using densely packed decima*l* encoding [2,3]. It is anticipated that, once the IEEE 754r Standard is finally approved, hardware support for decimal floating-point arithmetic on the processors will come into existence for financial, commercial, and Internet-based applications. Still, the major consideration while implementing BCD arithmetic will be to enhance its speed as much as possible.

Furthermore, researchers like Landauer have shown that for irreversible logic computations, each bit of information lost, generates $kT\log 2$ joules of heat energy, where k is Boltzmann's constant and T the absolute temperature at which computation is performed [4]. Bennett showed that $kT\ln 2$ energy dissipation would not occur, if a computation is carried out in a reversible way [5], since the amount of energy dissipated in a system bears a direct relationship to the number of bits erased during computation. Reversible circuits are those circuits that do not lose information and reversible computation in a system can be performed only when the system comprises of reversible gates. These circuits can generate unique output vector from each input vector, and vice versa, that is, there is a one-to-one mapping between input and output vectors. Classical logic gates are irreversible since input vector states cannot be uniquely reconstructed from the output vector states. There is a number of existing reversible gates such as Fredkin gate [6,7,8], Toffoli Gate (TG) [6, 7] and the New Gate (NG) [8]. As the Moore's law continues to hold, the processing power doubles every 18 months. The current irreversible technologies will dissipate a lot of heat and can reduce the life of the circuit. The reversible logic operations do not erase (lose) information and dissipate very less heat. Thus, reversible logic is likely to be in demand in high speed power aware circuits. Reversible circuits are of high interest in low-power CMOS design, optical computing, nanotechnology and quantum computing.

The most prominent application of reversible logic lies in quantum computers [10]. A quantum computer will be viewed as a quantum network (or a family of quantum networks) composed of quantum logic gates; each gate performing an elementary unitary operation on one, two or more two–state quantum systems called qubits. Each qubit represents an elementary unit of information; corresponding to the classical bit values 0 and 1. Any unitary operation is reversible and hence quantum networks effecting elementary arithmetic operations such as addition, multiplication and exponentiation cannot be directly deduced from their classical Boolean counterparts (classical logic gates such as AND or OR are clearly irreversible).Thus, quantum arithmetic must be built from reversible logical components.

One of the major constraints in reversible logic is to minimize the reversible gate used and garbage output produced. Firstly, this paper introduces, two novel BCD adder architectures termed CLA BCD (Carry look-ahead BCD) and CS BCD (Carry Skip BCD) adders respectively. The proposed BCD architectures are designed especially, to make them suitable for reversible logic synthesis. The first proposed CLA BCD adder is an improvement over the adder proposed in [13] and is modified to make it suitable for the reversible logic implementation. The second carry skip BCD adder is proposed to cater the need of carry skip adder in decimal arithmetic. Recently, a reversible conventional BCD adder was proposed in [11] using conventional reversible gates. Furthermore, this paper introduces a novel implementation of the BCD adders using a recently proposed TSG gate [12]. The TSG gate has the advantage that it can work singly as a reversible Full adder with only two garbage outputs. Thus, TSG gate is the most optimized gate for implementation of a reversible Full adder as far as known literature and our knowledge is concerned. It is being shown the proposed reversible implementation of the BCD adders is better than reversible BCD adder proposed in [11]. One new reversible gate, TS-3, is also proposed in this paper. Thus, an attempt has been tried to design faster BCD adders as well as to provide the platform for building decimal ALU of a Quantum CPU.

## 2. Conventional BCD Adder

A BCD adder is a circuit that adds two BCD digits in parallel and produces a sum digit also in BCD. Fig. 1 shows the conventional BCD adder. A BCD adder must also include the correction logic in its internal construction. The two decimal digits, together with the input carry, are first added in the top 4-bit binary adder to produce the binary sum. When the output carry is equal to zero, nothing is added to the binary sum. When it is equal to one, binary 0110 is added to the binary sum using another 4-bit binary adder (bottom binary adder). The output carry generated from the bottom binary adder is ignored, since it supplies information already available at the output carry terminal.

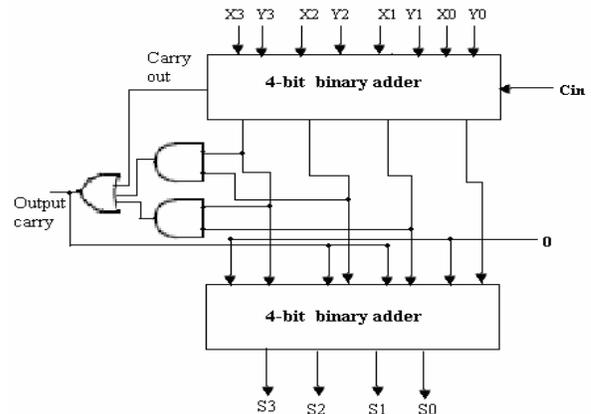

**Figure 1. Conventional BCD Adder**

## 3. Proposed Carry Look Ahead Adder

A Carry Look Ahead BCD Adder is proposed which is modification over the architecture proposed in [13] and is especially improved for making it suitable for CMOS as well as reversible logic implementation. In the proposed CLA BCD adder, OR gates used in the equations proposed in [13] are selectively chosen and replaced by XOR gates. One cannot replace randomly, the OR gates in equations of [13], thus a rigorous study has been done and OR gates in equations of [13] have been replaced at selective places. The functional verification of the proposed CLA BCD adder is done

in Verilog HDL using ModelSim simulator. Following are the advantages of using this approach

1. In the conventional CMOS logic, XOR gate can be designed, with less number of transistors compared to OR gate.

2. Recently, novel 2T-MUX architecture has been proposed in which the 2:1 MUX can be designed with only 2 transistors [14].

3. The advantage of the 2T MUX is that in addition to reduced transition activity and charge recycling capability, it has no direct connections to the power-supply nodes, leading to a noticeable reduction in short-current power consumption.

4. The XOR gate can be realized by using two 2T MUX as shown in the Fig.2 below

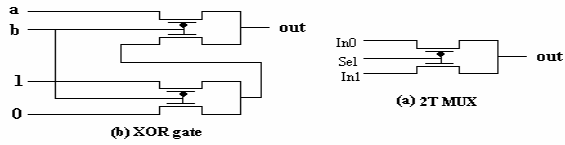

**Figure 2. (a) 2T MUX architecture (b) XOR using 2T MUX**

5. In the reversible logic, the multi-input XOR gate can be realized with less number of gates and garbage outputs compared to multi-input OR gate. For example, the equation $a \oplus b \oplus c$ can be realized with only one reversible gate and two garbage output compared to a+b+c ( here + refers an OR gate), which can be realized with two reversible gates and five garbage outputs. This advantage of XOR gate will become more dominant as the input size is increased beyond three.

Consider two BCD numbers a and b of 4 bits each, using the proposed approach, the functions used to generate Carry Look-Ahead BCD Adder are as follows

$$g[j] = a[j] \cdot b[j] \quad 0 \leq j \leq 3 \text{ "generate"}$$
$$p[j] = a[j] + b[j] \quad 0 \leq j \leq 3 \text{ "propagate"}$$
$$h[j] = a[j] \oplus b[j] \quad 0 \leq j \leq 3 \text{ "half-adder"}$$
$$m = g[3] \oplus (p[3] \cdot p[2]) \oplus (p[3] \cdot p[1]) \oplus (g[2] \cdot p[1])$$
$$n = p[3] \oplus g[2] \oplus (p[2] \cdot g[1])$$
$$C1 = g[0] + (p[0] \cdot Cin) \text{ "carry out of 1's position"}$$

$$S[0] = h[0] \oplus Cin$$
$$S[1] = ((h[1] \oplus m) \cdot C1) + (\sim (h[1] \oplus n) \cdot C1)$$
$$S[2] = (\sim p[2] \cdot g[1]) \oplus (\sim p[3] \cdot h[2] \cdot \sim p[1]) \oplus ((g[3] \oplus (h[2] \cdot h[1])) \cdot \sim C1)$$
$$\oplus (((\sim p[3] \cdot \sim p[2] \cdot p[1]) \oplus (g[2] \cdot g[1]) \oplus (p[3] \cdot p[2])) \cdot C1)$$
$$S[3] = ((\sim m \cdot n) \cdot \sim C1) \oplus (((g[3] \cdot \sim h[3]) \oplus (\sim h[3] \cdot h[2] \cdot h[1])) \cdot C1)$$
$$Cout = m + (n \cdot C1)$$

In the above equations, S[3], S[2], S[1], S[0] represents the sum bits produced by addition of BCD numbers, a and b and having a input carry Cin. The output carry produced by the CLA BCD adder is represented by Cout.

## 4. Proposed Carry Skip Adder

The proposed Carry Skip BCD Adder is being constructed in such a way that, the first full adder block consisting of 4 full adders can generate the output carry 'Cout' instantaneously, depending on the input signal and 'Cin', without waiting for the carry to be propagated in the ripple carry fashion. Fig.3 shows the proposed Carry Skip BCD adder. The working of the proposed Carry Skip BCD Adder (CS BCD Adder) can be explained as follows.

In the single bit full adder operation, if either input is a logical one, the cell will propagate the carry input to its carry output. Hence, the ith full adder carry input Ci, will propagate to its carry output, Ci+1, when Pi= Xi$\oplus$Yi where Xi and Yi represents the input signal to the ith full adder. In addition, the four full adders at the first level making a block can generate a "block" propagate signal 'P'. When 'P' is one, it will make the block carry input 'Cin', to propagate as the carry output 'Cout' of the BCD adder, without waiting for the actual propagation of carry, in the ripple carry fashion. An AND gate is used to generate a block propagate signal 'P'. Furthermore, depending on the value of 'Cout', appropriate action is taken. When it is equal to one, binary 0110 is added to the binary sum using another 4-bit binary adder (Second level or bottom binary adder). The output carry generated from the bottom binary adder is ignored, since it supplies information already available at the output carry terminal.

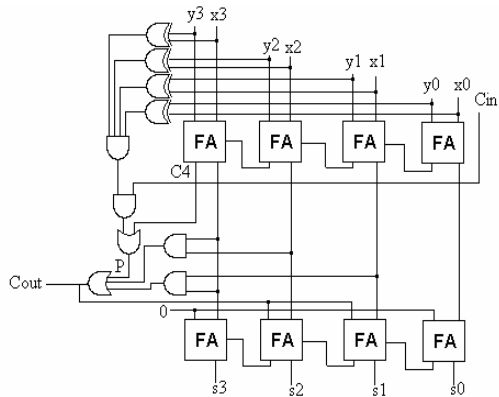

**Figure 3. Proposed Carry Skip BCD Adder**

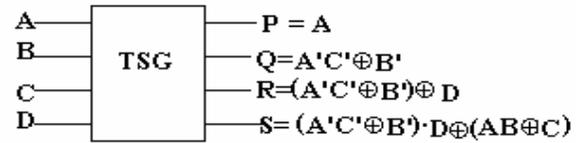

**Figure 4. Reversible 4 *4 TSG proposed in [12]**

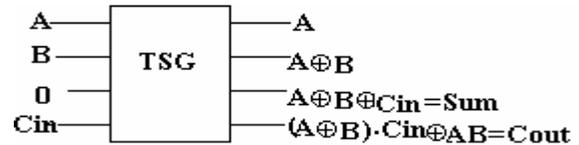

**Figure 5. TSG Gate working Singly as a Reversible Full Adder**

## 5. Reversible Logic Implementation of BCD Adders

In order to implement the reversible logic design of the conventional and the proposed BCD adders, some of the basic concepts of the reversible logic are discussed.

### 5.1. Basic Reversible Gates

There are number of a existing reversible gates in literature such as Fredkin gate [6,7,8], Toffoli Gate (TG) [6,7] and the New Gate (NG) [8]. Since, the major reversible gate used in designing the BCD adders is TSG gate, hence only the TSG gate is discussed in this section.

#### 5.1.1. TSG Gate

Recently, a 4 * 4 one through reversible gate called TS gate "TSG" is proposed [12]. The reversible TSG gate is shown in Fig. 4. The TSG gate can implement all Boolean functions. One of the prominent functionality of the TSG gate is that it can work singly as a reversible Full adder unit. Fig.5 shows the implementation of the TSG gate as a reversible Full adder. A number of reversible full adders were proposed in [15,16,17,18]. The full adder designed using TSG in Fig. 5 requires only one reversible gate (one TSG gate) and produces only two garbage outputs (Garbage output refers to the output that is not used for further computations. In other words, it is not used as a primary output or as an input to other gate.). It is shown that the full-adder design in Fig. 5 using TSG gate is better than the existing full-adder designs [12].

### 5.2 Novel Reversible TS-3 Gate

The proposed reversible TS-3 gate is a 3*3 two through gate as shown in Fig. 6. It can be verified from that in TS-3 gate, the input pattern corresponding to a particular output pattern can be uniquely determined.

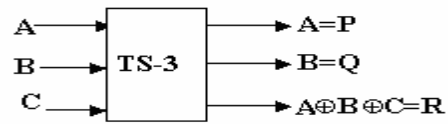

**Figure 6. Proposed 3 *3 Reversible TS-3 Gate**

The proposed TS-3 gate can be used to implement any Boolean function, since its third output can work as a three input XOR gate. Further, the proposed gate will be of great help in avoiding the fan out problem, since two its inputs are directly passed as output.

## 6. Reversible Logic Implementation of Conventional BCD Adder

Figure 7 shows the reversible implementation of the conventional BCD using the reversible TSG gates. For optimized implementation of the BCD adder; New Gate (NG) is also used for producing the best optimized BCD adder. Recently the reversible implementation of the conventional BCD adder is proposed in [11]. The BCD adder using the TSG gate is found to be much better than the architecture proposed in [11]; both in terms of number of reversible gates and garbage outputs produced. The proposed BCD adder architecture in Fig. 7 using TSG gates uses only 11 reversible gates and produces only 22 garbage outputs, compared to 23 reversible gates and 22 garbage outputs implementation of [11]. Table 1 shows the result which compares the proposed reversible BCD adder using TSG gate with the BCD adder proposed in [11].

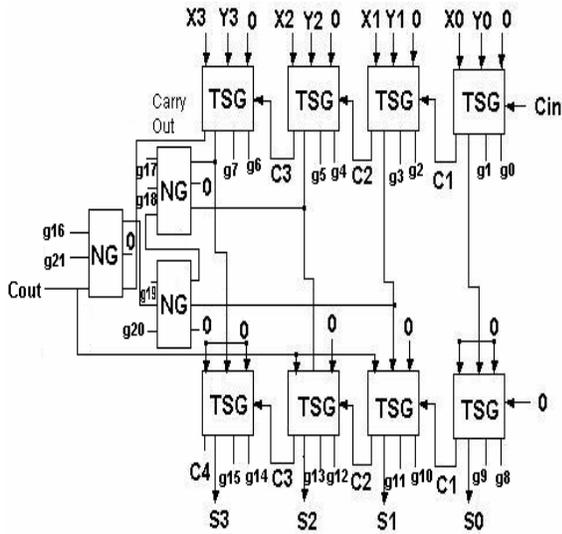

**Figure 7. Reversible Logic Implementation of the Conventional BCD Adder Using TSG and NG Gates**

## 7. Reversible Logic Implementation of Carry Skip BCD Adder

The proposed Carry Skip Adder is further been modified to make it more suitable for reversible logic implementation. Fig. 8 shows the block diagram of the carry skip adder block constructed with TSG gates and Fredkin gates (F). The three Fredkins in the middle of Fig. 8 are used to perform the AND4 operation. This will generate the block propagate signal 'P'. The single FG in the right side of Fig. 8 performs the AND-OR function while in conventional logic, the carry skip block in Fig.3 uses the AND-OR gate combination to create the carry skip logic.

In the proposed carry Skip adder, the FG propagates the block's carry input 'Cin' to the next block if the block propagate signal 'P' is one ; otherwise, the most significant full adder carry 'C4' is propagated to the next block. The traditional carry skip AND-OR logic in Fig.3 and the carry skip logic in Fig. 8 do not have equivalent truth tables, but it must be noted that the Fredkin carry skip logic more faithfully adheres to the spirit of carry skip addition, by propagating the correct value of 'Cin' to 'Cout'. In the AND-OR logic combination in Fig. 3 generation of 'Cout' will be postponed until 'C4' is resolved, while in the proposed reversible architecture, Fredkin carry skip logic passes 'Cin' to 'Cout' whenever 'P'=1, regardless of the value of 'C4', thus time saving is significant. Furthermore, in the proposed carry skip adder of Fig.3, it has been observed that the 3 input OR gate, used for generating the 'Cout' can be replaced by three input XOR gate. Thus, the novel proposed 'TS-3' gate helps in the realization of 3 inputs XOR gate, with bare minimum of one reversible gate. Additionally, it can also help in avoiding the problem of fan out, as two of the inputs are directly passed as outputs. The numbers of reversible gates used in the proposed reversible logic implementation of the Carry Skip BCD adder are 15 and the garbage outputs produced are 27.

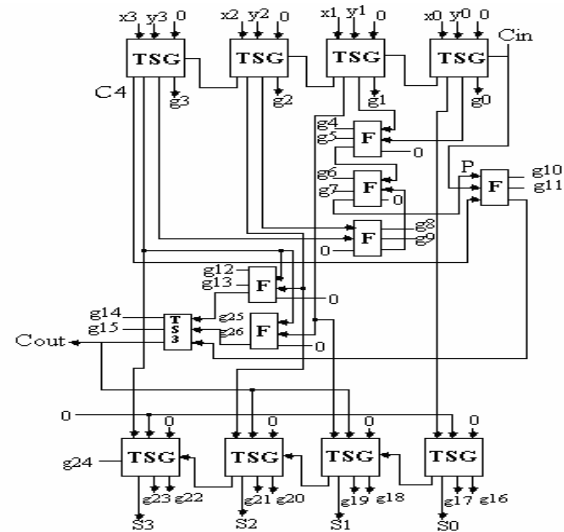

**Figure 8. Reversible Logic Implementation of the Proposed Carry Skip BCD Adder**

The proposed reversible implementation of the Carry Skip BCD adder is still better, compared to the reversible logic implementation of the conventional BCD adder proposed in [11], in terms of reversible gates used. A comparative analysis of the BCD adders is shown in Table 1. Furthermore, the time saving in the proposed reversible carry skip BCD adder is significant compared to [11]. This is due to carry skipping and this feature was not available in the reversible BCD adder of [11], since it was a mere conventional BCD adder.

**Table 1. Comparative Analysis of the Reversible BCD Adders**

|  | Number of Reversible Gates Used | Number of Garbage Outputs Produced |
|---|---|---|
| Existing Reversible BCD Adder[11] | 23 | 22 |
| Proposed Reversible Conventional BCD Adder | 11 | 22 |
| Proposed Reversible Carry Skip BCD Adder | 15 | 27 |

## 8. Conclusions

The focus of this paper is the IEEE 754r which is the ongoing revision to the IEEE 754 floating point standard considering decimal arithmetic. Thus, this paper proposes two novel designs of BCD adder called Carry Look Ahead and Carry Skip BCD Adders respectively. The architectures are specially designed to make them suitable for reversible logic implementation. A new reversible gate TS-3 is also proposed in this paper and its optimized application is done at the appropriate place. The reversible logic synthesis (design) is being done, for both the conventional as well as the proposed carry skip BCD adder. It has been shown that the proposed reversible designs of the BCD adders are much better compared to its counterpart proposed in [11], both in terms of number of reversible gates and garbage outputs produced. The proposed circuit can be used for designing large reversible systems which is the necessary requirement of quantum computers, since quantum computers must be built from reversible components. Thus, the paper provides the initial threshold to build more complex systems which can execute more complicated operations. The reversible circuits designed and proposed here, form the basis of the BCD ALU of a primitive quantum CPU.